\documentclass[pra,aps,twocolumn,reprint,floatfix,amsmath,amssymb,superscriptaddress]{revtex4-1}
\usepackage{hyperref}
\usepackage{color}
\usepackage{graphicx}
\usepackage{braket}

\newcommand{\vc}[1]{\ensuremath{#1}}

\begin{document}

\title{Photon-mediated qubit interactions in 1D discrete and continous models}
\author{Guillermo D{\'\i}az-Camacho}
\affiliation{Instituto de F{\'\i}sica Fundamental IFF-CSIC, Calle Serrano 113b, Madrid E-28039, Spain}
\author{Diego Porras}
\affiliation{Department of Physics and Astronomy, University of Sussex, Falmer, Brighton BN19QH, UK}
\author{Juan Jos\'e Garc{\'\i}a-Ripoll}
\affiliation{Instituto de F{\'\i}sica Fundamental IFF-CSIC, Calle Serrano 113b, Madrid E-28039, Spain}
\email{jj.garcia.ripoll@csic.es}

\begin{abstract}
  In this work we study numerically and analytically the interaction of two qubits in a one-dimensional waveguide, as mediated by the photons that propagate through the guide. We develop strategies to assert the Markovianity of the problem, the effective qubit-qubit interactions and their individual and collective spontaneous emission. We prove the existence of collective Lamb-shifts that affect the qubit-qubit interactions and the dependency of coherent and incoherent interactions on the qubit separation. We also develop the scattering theory associated to these models and prove single photon spectroscopy does probes the renormalized resonances of the single- and multi-qubit models, in sharp contrast with earlier toy models where individual and collective Lamb shifts cancel.
\end{abstract}

\maketitle

\section{Introduction}

Technological progress in the fields of superconducting quantum circuits\ \cite{astafiev10} and nanophotonics\ \cite{lodahl13} are opening the door to new quantum technologies based on propagating photons and few-level systems, such as few-photon transistors\ \cite{chang07,hoi11}, non-classical states of the radiation field\ \cite{hoi12,eichler12} or new photodetector schemes\ \cite{romero09a,romero09b}. Of particular fundamental interest are also proposals to use these low-dimensional photonic systems to mediate interactions between qubits\ \cite{gonzalez-tudela13,gonzalez-tudela11,martin-cano11}, for which there are already initial implementations in the superconducting world\ \cite{vanloo13,lalumiere13}, and which could potentially have applications in the quantum information world, as well as in quantum simulation and the study of quantum phase transitions.

At the same time that this technology evolves, we need to further develop the theoretical tools that study the light-matter and light-mediated matter interactions. While there exist approximation methods for some of the optical technologies mentioned above, based on single-photon single-qubit effective boundary conditions\ \cite{shen05a,shen05b}, input-output theory\ \cite{lalumiere13,caneva15} or scattering theory\ \cite{shi11,laakso14}, the degree of control of some methods or their computational generality for any number of photons is an open problem.

Matrix Product State (MPS) methods are one novel tool that solves some of the limitations of earlier methods regarding photon numbers and interaction strengths to describe the full light-matter wavefunction and its time evolution. These methods rely on a \textit{discretization} of the photonic degrees of freedom either in frequency or in position space, and seem to provide qualitatively and even quantitatively accurate results for problems of few photon scattering\ \cite{peropadre13,sanchez-burillo14a,sanchez-burillo14b}. Two questions open: how can we validate these models and discretizations and how do they compare to the usual techniques, i.e. master equations and Markovian approximations.

This work dwells into these problems by studying a very precise feature of the numerical approximations mentioned above: the discretization of the photonic degrees of freedom. We study the effective interaction of two qubits in the weak- and strong-coupling regime, both within the Rotating Wave approximation and with Matrix Product States, for a variety of underlying microscopic models of the photonic degrees of freedom. Combining theoretical estimates based on the Wigner-Weisskopf theory with exact analytical studies of the single-excitation or Rotating Wave Approximation (RWA) limit we prove that the Markovian approximation is only well justified for qubits inside their mutual light-cones and develop accurate numerical procedures to extract the effective parameters of the qubit-qubit interaction and collective decay. These parameters agree qualitatively with earlier works based on the resonant-dipole approximation\ \cite{gonzalez-tudela13}. However, we also find collective effects (Lamb-shifts) which strongly depend on the cut-offs and the microscopic details of the interaction, and which renormalize the qubit frequencies, their interactions and the effective individual and collective dissipation. The renormalization effects can be of measurable magnitude, increasing with the qubit-qubit separation, and we suggest how recent experiments with superconducting circuits could probe those features.

There are two reasons why one could doubt of these non-universal renormalization effect: such effects have not been found in theoretical studies of single-photon scattering\ \cite{shen05a,shen05b} and we have derived them using RWA equations. We dispel the first criticism studing the Lippmann-Schwinger scattering formalism and proving that its resonances are determined by the same renormalized parameters of the theoretical model. We also study existing single-photon scattering literature and conclude that the "toy"-models used to represent the photons are often too simple and have zero Lamb shifts. Finally, we report on quasi-exact numerical simulations of the full spin-boson model including counter-rotating terms, for exactly the same parameters of the RWA simulations. These MPS simulations confirm the RWA results for coupling strengths below five percent of the qubit gap, $g/\Delta<5\%$, which is the current operating regime for circuit-QED in open transmission lines\ \cite{vanloo13,hoi12,hoi11,astafiev10}.

The structure of this manuscript is as follows. In Sect.\ \ref{sec:models} we introduce a simplified light-matter interaction model for two qubits that talk to a one-dimensional photonic medium. We derive the evolution equations for the qubits in the single-excitation limit and obtain the Markovian approximation and resulting master equation with care. This derivation pays great attention to the infrarred and ultraviolet cut-offs of the problem, leading us to conclude that individual and collective Lamb-shift influence the qubit frequencies, and the spontaneous and collective emission rates of the qubits beyond earlier predictions. In Sect.\ \ref{sec:ls-theory} we relate the parameters of the Markovian model with the Lippmann-Schiwnger scattering states and dicuss why earlier scattering studies did not show the renormalization effects. While the theoretical results so far were based on a few solvable models, in Sects.\ \ref{sec:gapless} and \ref{sec:band} we return to physically realistic models of waveguides and of photonic crystals. We confirm the role of the cut-off frequency in the single-qubit renormalization and in the two-qubit interactions in both cases. In Sect.\ \ref{sec:mps} we introduce numerical simulations based on Matrix Product States and confirm that they agree with the theoretical predictions in the strong-coupling limit. Finally in Sect.\ \ref{sec:summary} we summarize the main consequences of this study and discuss possible experimental studies of these non-universal effects. 

\section{Theoretical models}
\label{sec:models}

\subsection{Spin-boson model}

We wish to study the interaction of one or more two-level systems with a photonic medium. For that we rely on the spin-boson model, which is a convenient description for the type of implementations where this problem has some interest ---namely superconducting circuits and nanophotonics. The model reads
\begin{equation}
  H = \sum_i\frac{\Delta_i}{2} \sigma_i^z + \sum_{k} \omega_k a^\dag_{k} a_{k}
  + \sum_{k,i} \sigma^x_i (g_{ik}  a_{k} + g_{ik}^{*}  a^\dag_{k}).
  \label{eq:spin-boson}
\end{equation}
The light-matter coupling typically adopts the form
\begin{equation}
  g_{ik} = g \sqrt{\frac{\omega_k}{2}} \times\frac{e^{ikx_i}}{\sqrt{L}},
  \label{eq:coupling}
\end{equation}
where $\omega_k$ is the dispersion relation, $x_i$ is the position of the $i$-th qubit in real space, $L$ is the physical length of the one-dimensional medium and $g$ is a parameter measuring the coupling strength.

As discussed below, the model\ \eqref{eq:spin-boson} adopts different dispersion relations and coupling strengths depending on the microscopic description of the underlying photonic medium. For computational reasons, in any of these descriptions we will have to impose both a finite medium size, $L$, and a spatial discretization $\Delta{x} = L/N$. This leads to a discretization of the momenta
\begin{equation}
  k \in \frac{2\pi}{L}\times\{0,\pm 1,\pm 2\ldots \pm \frac{N}{2}\}.
\end{equation}

In either case we may introduce the spectral function
\begin{equation}
  J(\omega) = 2\pi\sum_{k} \left|g_{ik}\right|^2 \delta(\omega - \omega_{k}).
\end{equation}
Using the fact that momenta are equispaced, and that the frequency spacing $\mathrm{d}\omega = \omega'(k) \mathrm{d}k = \mathrm{d}k/\rho(\omega)$ is inversely proportional to the density of states, $\rho(\omega)$, we may write
\begin{align}
  J(\omega) &= 2 \times 2\pi\int_0^{\omega_c} g^2 \omega_{k} \delta(\omega - \omega_{k}) \rho(\omega_k)\mathrm{d}\omega_k\\
  &= 4\pi g^2 \omega \rho(\omega).\nonumber
\end{align}
If we assume the RWA and study the spontaneous emission of a two-level system interacting with this waveguide, we will obtain that the spontaneous emission rate is $\gamma/2 = J(\omega)/2 = g^2\omega\rho(\omega)$, independent of the microscopic details of the discretization we used.

\subsection{Microscopic models}
\label{sec:microscopic}

We will typically consider two types of problems: gapless photons and a photonic crystal with a finite bandwidth. The first case corresponds to a chain of coupled oscillators and it models the propagation of optical photons or microwave photons in a waveguide\ \cite{peropadre13}. The dispersion relation has the form
\begin{equation}
  \omega_k = \omega_c\sqrt{2 - 2\cos(k\Delta x)},\;
  g_{sk} = g \sqrt{\frac{\omega_k}{2L}}e^{ikx_s}
  \label{eq:w-gapless}
\end{equation}
with a cut-off $\omega_c = 1/\Delta{x}$ that depends on the discretization. With this choice of units, the dispersion relation becomes linear at low momenta, $\omega_k \simeq v |k|$, with unit speed of light, $v=1$.

The other possible problem that we will study is a chain of coupled cavities. In this case the dispersion relation has a finite bandwidth
\begin{equation}
  \omega_k = \omega_0 - J\cos(k\Delta x),\;
  g_{sk} = g \sqrt{\frac{\omega_k}{2L}}e^{ikx_s}
  \label{eq:w-band}
\end{equation}
and the discretization step plays no role, so that we can take $\Delta{x}:=1$.

Finally, in addition to these models with a finite number of modes, we will consider a continuum limit in frequency space that has $L\to\infty$ and introduces the cut-off in the coupling strength, not on the dispersion relation
\begin{equation}
  \omega_k = |k|,\;
  g_{sk} = g e^{-\omega_k/\omega_c}\sqrt{\frac{\omega_k}{2L}}e^{ikx_s}.
  \label{eq:w-leggett}
\end{equation}
This exponential cut-off is usually employed in studies of the spin-boson model\ \cite{leggett87} because it allows efficient computation of memory functions and analytical predictions.

In addition to these realistic models, Sect.\ \ref{sec:status-quo} analyzes other models that have become standard in the literature, such as the linearized model by Shen and Fan\ \cite{shen05a,shen05b} or coupled cavities models\ \cite{zhou2008,longo2010}, where the coupling strength is approximated as frequency independent, leading to $J(\omega)$ approximately constant.

\subsection{RWA and single-excitation equations}

In the low-energy and small-coupling limit, in which $g$ is small compared with the qubit and photon frequencies\ \cite{peropadre13}, we can simplify the spin-boson model adopting the so called rotating-wave approximation, in which the counterrotating terms $\sigma^+_ia^\dag_k+\mathrm{H.c.}$ are neglected. In the RWA limit the number of excitations in the system is conserved and we can write down variational wavefunctions in which either one of the qubits is excited, or a photon is propagated. We write this as
\begin{equation}
  \label{eq:single-excitation}
  \ket{\psi} = \left[\sum_i c_i \sigma^+_i + \sum_k \psi_k\right]
  \ket{\Omega},
\end{equation}
where $\ket{\Omega}=\otimes_i\ket{g_i}\otimes_k\ket{0_k}$ is the vacuum in the qubit and photon spaces.

The evolution equations for the state read
\begin{align}
\label{eq:ode-exact}
i\partial_t c_s &= \Delta_s c_s + \sum_k g_{ik} \psi_k,\\
i\partial_t \psi_k &= \omega_k \psi_k + \sum_s g_{sk}^* c_{s}.\nonumber
\end{align}
We integrate formally the photons
and plug this expression into the qubits
\begin{equation}
i\partial_t c_s = \Delta_s c_s + a_s(t) + \sum_{s'}b_{ss'}(t).
\label{eq:integro-differential}
\end{equation}
Each qubit has an input signal that either comes from the photonic field, or from a memory function that concerns the dynamics of all of the qubits in the past
\begin{align}
a_s(t) &= \sum_k g_{sk} e^{-i\omega_k t} \psi_k(0),\\
b_{ss'}(t) &= -i \int_0^t K_{ss'}(t-\tau)c_{s'}(\tau)\mathrm{d}\tau  
\end{align}
with the kernel
\begin{align}
  K_{ss'}(t) & = \int \frac{J(\omega_k)}{4\pi} e^{-i\omega_k(t-\tau)}
  \left[e^{-ik (x_s-x_{s'})} + \mathrm{c.c.}\right]\mathrm{d}\omega_k\nonumber \\
  & = K(t; d_{ss'}) + K(t; -d_{ss'}).
\end{align}
For a linear dispersion relation, $\omega_k=v|k|$, the previous kernel adopts the form $K(t+d_{ss'})+K(t-d_{ss'})$, implying that the separation between qubits induces a delay in the mutual interaction.

\section{Master equations}
\label{sec:markovian}

\subsection{Single-qubit master equation limit}

Equation\ \eqref{eq:ode-exact} can be manipulated to study the dynamics of the qubit and the field under further approximations. The most common one is the Markov approximation in which we assume that the memory of the field decays much faster than the time it takes for the qubit to evolve.

If we have a single qubit, we will be concerned with the evolution of the single memory function $K_{ss}(t)$. We assume a-priori, that $c(t)$ evolves as $\exp(-i\Delta't)w(t)$, with some renormalized frequency $\Delta'$ that is close to the original qubit frequency $\Delta$, and a slow function $w(t)$:
\begin{align}
  b_{ss}(t) &= -i \int_0^t K_{ss}(t-\tau) e^{-i\Delta'\tau}w(\tau)\mathrm{d}\tau\\
  &\simeq -i w_s(t)e^{-i\Delta_s't}
  \int_{-\infty}^t K_{ss}(t-\tau) e^{i\Delta_s'(t-\tau)}\mathrm{d}\tau\nonumber
\end{align}
This leads to a memoryless equation for one qubit
\begin{equation}
  i\partial_t c_s = \left(\Delta_s' - i \frac{\gamma_s}{2}\right) c_s,
\end{equation}
replacing the memory function with the constants
\begin{equation}
  \delta_s - i\frac{\gamma_s}{2} =
  \int_0^\infty 2K(\tau;0)e^{i\Delta_s'\tau}\mathrm{d}\tau.
  \label{eq:1qb-parameters}
\end{equation}
Using the identity
\begin{equation}
  \int_0^\infty e^{-i(\omega-\Delta') \tau}\mathrm{d}\tau = \left[\pi\delta(\omega-\Delta')
    -i\mathrm{PV}\left(\frac{1}{\Delta' -\omega}\right)\right],
\end{equation}
we obtain the correction to the qubit frequency and the decay rate as solutions of the self-consistency equation
\begin{align}
  \label{eq:self-consistent}
  \delta_s &= \Delta_s' - \Delta_s =
  \frac{1}{\pi}\mathrm{PV} \int \frac{J(\omega)}{(\Delta_s'-\omega)}\mathrm{d}\omega,\\
  \gamma_s & = J(\Delta_s')
\end{align}

\begin{figure}
  \centering
  \includegraphics[width=0.8\linewidth]{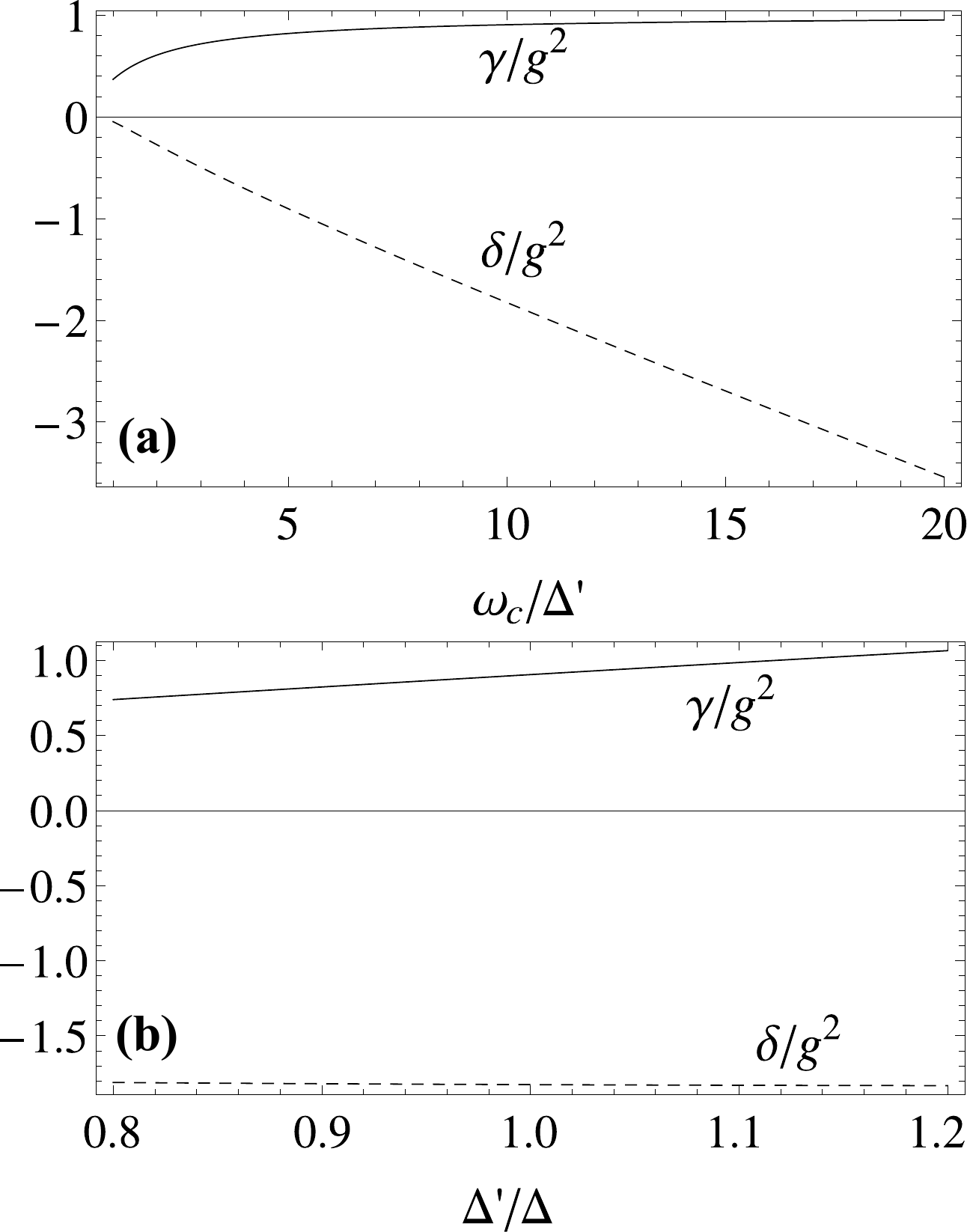}
  \caption{Parameters for the evolution of the two-level system coupled to the 1D transmission line as a function of the cut-off, computed with Eq.\ \eqref{eq:1qb-parameters} for model\ \eqref{eq:w-leggett}. In (a) we fix $\Delta=1$ and try different cut-offs. In (b) we fix the cut-off $\omega_c=10\Delta$ and change $\Delta'$. It is evident from this plot that since $|\Delta-\Delta'|\simeq g^2\ll 1$, there is no big difference between using $\Delta$ and $\Delta'$ in Eq.\ \eqref{eq:self-consistent}. \label{fig:1qb-analytics}}
\end{figure}

In Fig.\ \ref{fig:1qb-analytics}a we summarize the outcome of these computations for the exponential cut-off\ \eqref{eq:w-leggett}, for which we can compute the kernel exactly
\begin{align}
  \gamma &= g^2\Delta' e^{-\Delta'/\omega_c},
  \label{eq:simple-renorm}\\
  \delta &= -g^2\frac{1}{2\pi}\left[\omega_c - e^{-\Delta'/\omega_c}\mathrm{Ei}(\Delta'/\omega_c)\right],\nonumber
\end{align}
where $Ei(x)$ is the exponential integral, a negligible term when $\omega_c\to\infty$. Note how the renormalization of the qubit frequency $\delta$ may become significantly larger than the decay rate itself, diverging as the cut-off increases. However, even if this non-universal effect is comparatively large, the ratio $\delta/\Delta$ is still small and Fig.\ \ref{fig:1qb-analytics}b shows that we can solve Eq.\ \eqref{eq:self-consistent} in a simpler, approximate way, using $\Delta$ instead of $\Delta'$ in the integral, and replacing the resulting correction $\Delta'$ in the expression for $\gamma$
\begin{align}
  \Delta_s' &\simeq
  \frac{1}{\pi}\mathrm{PV} \int \frac{J(\omega)}{(\Delta_s-\omega)}\mathrm{d}\omega,\label{eq:simple-renorm2}\\
  \gamma_s & = J(\Delta_s').\nonumber
\end{align}

\subsection{Two-qubit master equation limit}

The same procedure can be repeated for two or more qubits. In this case the outcome should read
\begin{equation}
  i \partial_t c_s = \left(\Delta_s' - i \frac{\gamma_s}{2}\right) c_s +
  \sum_{s' \neq s}\left(g_{ss'} - i \frac{\gamma_{ss'}}{2}\right)c_{s'},
  \label{eq:nqb}
\end{equation}
with the additional constants
\begin{equation}
  g_{ss'}-i\frac{\gamma_{ss'}}{2}=\int_0^\infty K_{ss'}(\tau) e^{i\Delta_{s'}'\tau}
  \mathrm{d}\tau.
  \label{eq:nqb2}
\end{equation}
There are several caveats on Eq.\ \eqref{eq:nqb}, though. The first one is that this equation now is only valid for times $t > T_{ss'} = |d_{ss'}|/v$, larger than the time it takes a photon emitted by qubit $s$ to arrive to qubit $s'$. Before that time, as we will see below, the qubits evolve almost independent from each other, in an approximate causality. 

There are also some subtleties when we try to estimate the parameters $g_{ss'}$ and $\gamma_{ss'}$. Firstly, Eq.\ \eqref{eq:nqb2} neglects the effect of the collective coupling $g_{ss'}$ in the renormalization of the single-qubit frequencies $\Delta_s$. Secondly, $g_{ss'}$ is related to the principal value part of the spectral function. We cannot compute this in closed form because the RWA precludes the use of the Kamers-Kronning relation, but assuming that non-RWA terms are negligible we can approximate $g_{ss'}$ by the expression that results from the full model (See Suppl.\ Matt. in\ \cite{gonzalez-tudela13}). Based on these considerations, an approximate set of constants for this problem that is often used in the literature, combines Eq.\ \eqref{eq:self-consistent} with the resonant dipole approximation
\begin{align}
\label{eq:collective}
  \gamma_{ss'} &\simeq \gamma_s' \cos(k_sd_{ss'}),
  \quad d_{ss'}\gg v/\omega_c\\
  g_{ss'} &\simeq \frac{\gamma_s'}{2}\sin(k_s|d_{ss'}|),\nonumber
\end{align}
where $k_s\ge0$ is the positive momentum associated to $\Delta_s'$. An important prediction of this formula is that, as long as it is valid, \textit{dipole-dipole interactions can be suppressed} by placing quantum emitters at distances such that $k_s |d_{s,s'}| = m \pi$. This condition leads to the implementation of homogeneous superadiant models \cite{Chang12,gonzalez-tudela13}.

When the qubit have identical frequencies, $\Delta_1=\Delta_2=\Delta$, we can develop a more accurate approach to the Markovian limit that takes into account self-consistent renormalization. In this case the exact problem decouples into two equations, with variables $c_\pm = \frac{1}{\sqrt{2}}(c_1\pm c_2)$
\begin{equation}
  i\partial c_\pm = \Delta c_\pm
  -i \int_0^t K_\pm (t-\tau)c_\pm(\tau)\mathrm{d}\tau
  \simeq (\Delta_\pm-i\gamma_\pm)c_\pm.
\end{equation}
with $K_\pm(t)=2K(t;0)\pm[K(t;d)+K(t;-d)]$ and $d$ the distance between the two qubits. From these equations we recover
\begin{align}
  &\Delta'= \frac{1}{2}(\Delta_++\Delta_-),
  &\gamma= \frac{1}{2}(\gamma_++\gamma_-),\label{eq:2qb-parameters}\\
  &g_{12} = \frac{1}{2}(\Delta_+-\Delta_-),
  &\gamma_{12} = \frac{1}{2}(\gamma_+-\gamma_-)\nonumber
\end{align}
in a more correct than the derivation above. Let us remark the need to compute $\Delta_\pm$ separately, using either\ \eqref{eq:self-consistent} or\ \eqref{eq:simple-renorm2}. This gives rise to collective renormalization effects that depend on $g_{12}$ and these effects influence the values of $\gamma$ and $\gamma_{12}$, in contrast with Eq.\ \eqref{eq:collective}, where the spontaneous emission parameters only depended on $\Delta'$.

\begin{figure}
\includegraphics[width=0.9\linewidth]{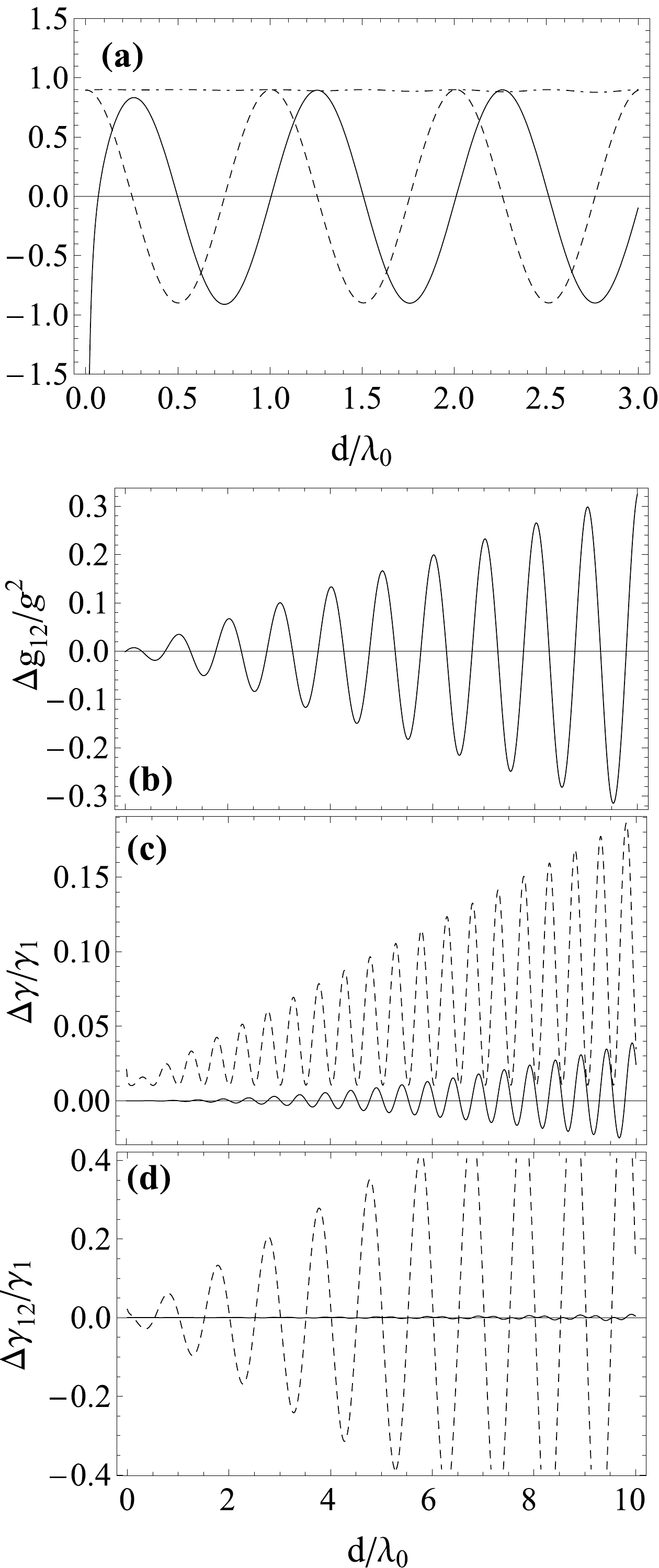}
  \caption{(a) Parameters for the evolution of the two qubits coupled to the 1D transmission line as a function of the distance between qubits, measured in units of $\lambda_0$, the qubit wavelength. We plot $\gamma/2g^2$ (dash-dot), $\gamma_{12}/2g^2$ (dashed) and $g_{12}/g^2$ (solid), computed using\ \eqref{eq:2qb-parameters}, using a self-consistent formula for $\Delta_{\pm}$. (b-d) Errors in the parameters $g_{12}$, $\gamma_{12}$ and $\gamma$ made by using no Lamb-shift renormalization (dashed) or by using the simplified renormalization scheme\ \eqref{eq:simple-renorm2} (solid). Note that the error is amplified with respect to the single-qubit case because already $\gamma\sim 2\gamma_1$. In all simulations $\omega_c=10\Delta$ and $g=0.04\Delta$.\label{fig:2qb-analytics}}
\end{figure}

We can particularize earlier expressions for the model\ \eqref{eq:w-leggett}. Keeping the lowest order terms in the coupling, $g$, and dropping the corrections from using Eq.\ \eqref{eq:2qb-parameters}, the Markovian interaction reads,
\begin{align}
  \Delta' &=
  \Delta - g^2\frac{1}{\pi}[\omega_c + e^{-\Delta/\omega_c}Ei(\Delta/\omega_c)],\label{eq:2qb-formulas-a}\\
  g_{12} &= -\frac{g^2}{2\pi}\frac{2\omega_c}{1+d^2\omega_c} \\
  &+ \frac{g^2\Delta e^{-\Delta/\omega_c}}{2} \left( \sin(d\Delta) +  f(d\Delta) \right),\nonumber\\
  \gamma &= g^2e^{-\Delta/\omega_c}\Delta + \mathcal{O}(g^4\omega_c)\\
  \gamma_{12} &= g^2\Delta e^{-\Delta/\omega_c}\cos(d\Delta) + \mathcal{O}(g^4\omega_c),\label{eq:2qb-formulas-b}
\end{align}
where we introduce $f(x)=- \mathrm{Re}\left\{e^{i x}E_1[ix + \Delta/\omega_c)]\right\}$. This result agrees qualitatively with the approximation in Eq.\ \eqref{eq:collective} in the limit $d\Delta \gg 1$, where interaction becomes approximately periodic. The qubit interaction, however, diverges at short distances and becomes of order $\delta=-g^2\omega_c$. This happens at a distance that is inversely proportional to the cut-off, which is the quantity that determines our \textit{spatial resolution}. Moreover, as we will see below, the renormalization of the frequencies introduces large discrepancies with\ \eqref{eq:collective}.

In Fig.\ \ref{fig:2qb-analytics}a we plot the constants that result from an exact self-consistent evaluation of the Markovian parameters through a direct evaluation of the kernel integral for the model\ \eqref{eq:w-leggett}. The inclusion of the self-consistent corrections does not change the qualitative shape of $g_{12}$ and $\gamma_{12}$, but because $\Delta_+$ is not identical to $\Delta_-$, there are beatings that appear in $\gamma$ and which amplify with the distance and the cut-off. In order to investigate this further, Fig.\ \ref{fig:2qb-analytics} we have taken the difference between the full, self-consistent solution in Fig.\ \ref{fig:2qb-analytics}a, and a solution without Lamb shift corrections (dashed) and where these corrections are introduced only in $\gamma$ (solid), as in Eq.\ \eqref{eq:simple-renorm2}. These differences are shown in Figs.\ \ref{fig:2qb-analytics}b-d for $g_{12}$, $\gamma$ and $\gamma_{12}$. Unlike for a single qubit, there are sizeable discrepancies even in the coherent part of the evolution $g_{12}$. Note also how $\gamma$ differs enormously from the trivial, distance independent prediction, and presents those beatings mentioned before. Finally, $\gamma_{12}$ also differs from the unrenormalized case, but it seems that already Eq.\ \ref{eq:simple-renorm2} provides a good enough approximation.

\section{Lippmann-Schwinger scattering theory}
\label{sec:ls-theory}

So far we have derived effective models for the nonequilibrium dynamics of the qubits themselves, tracing out the photons. We will now prove that the resonances of this effective master equation may be probed using single-photon spectroscopy to excite the internal transitions of the effective spin-spin interaction and accessing $\gamma$, $\gamma_{12}$ and $g_{12}$ through the changes in state of the photons that interact with the qubits themselves.

Given the constraints of our model, which is developed in frequency space, we cannot resort to the direct computation of scattering matrices\ \cite{shen05a,shen05b}. Instead we will use the resolvent method, introducing the basic idea in Sect.\ \ref{sec:resolvent} and applying it to single- and two-qubit processes in subsequent texts.

\subsection{Scattering states}
\label{sec:resolvent}

We study the propagation of waves when they face an impurity or potential represented by a local interaction $V$ following the operator formalism that was set up by Lipmman and Schwinger and is nicely summarized in Ref.\ \cite{ballentine14}. The original problem admitted a continuum of solutions labeled by some momentum $\vc{k}$ and energy $E_{\vc{k}}$
\begin{equation}
  (H_0 - E_{\vc{k}}) \ket{\Phi_{\vc{k}}} = 0.
\end{equation}
The new solutions are expected to assimilate to the reflected beams out of an incoming wave $\ket{\Phi_{\vc{k}}}$. Because far away from the perturbation they must contain such plane wave, and at those distances the dynamics is dominated by $H_0$, these solutions must have the same energy $E$ but satisfy a different equation
\begin{equation}
  (H - E_{\vc{k}}) \ket{\Psi_{\vc{k}}} = (H_0 + V - E_{\vc{k}}) \ket{\Psi_{\vc{k}}} = 0.
\end{equation}
For simplicity we are also going to drop the $\vc{k}$ label unless explicitly needed.

The Lippmann-Schwinger (LS) formalism starts by defining the resolvents of the free and interacting problems
\begin{align}
  G_0(E) = (E - H_0)^{-1},\;  G(E) = (E - H)^{-1}.
\end{align}
We also introduce the quantity $E^+=E+i\varepsilon$, modifying the scattering problem to
\begin{equation}
  \label{eq:H}
  (E^+ - H) \ket{\Psi} = 0.
\end{equation}
where $\varepsilon>0$ will be eventually taken to the limit $\varepsilon\to 0^+$. The sign of this small perturbation is relevant to distinguish the direct problem ($\Phi$ represents the incoming wave), from the inverse problem ($\Phi$ is the time reversed of the scattered wave).

From \eqref{eq:H} we construct a self-consistent solution
\begin{equation}
  \label{eq:LS}
  \ket{\Psi} = \ket{\Phi} + G_0(E^+) V \ket{\Psi} = \sum_{n=0}[G_0(E^+)V]^n\ket{\Phi}
\end{equation}
If we apply the operator $(E^+ - H_0)$ on this equation and use the fact that $\ket{\Phi}$ belongs to its kernel, we recover\ \eqref{eq:H}. The $\ket{\Phi}$ provides the appropriate incoming boundary condition, while the second term represents all the scattered waves.

The LS equation\ \eqref{eq:LS} can be solved self-consistently. The first order truncation of the series is the so called Born approximation
\begin{equation}
  \label{eq:Born}
  \ket{\Psi} = \left[1 + G_0(E^+) V\right] \ket{\Phi}.
\end{equation}
This is a convenient approximation that is reliable under some limits that include a tight localization in time and space of the scatterer, so that we do not need to consider multiple absorptions and reemissions of the scattered particle --- a sort of RWA. However, this is not the only solution to the problem. If we are able to invert the full Hamiltonian approximately we can construct the solution $\ket{\Phi}$ as
\begin{equation}
  \label{eq:LS-full}
  \ket{\Psi} = \left[1 + G(E^+)V\right]\ket{\Phi} =: \ket{\Phi} + \ket{\boldsymbol\psi},
\end{equation}
where $\ket{\boldsymbol\psi}$ is the outgoing or scattered state. To prove this expression we use\ \eqref{eq:LS} and the relations $[1 + G(z)V][1 - G_0(z)V] = [1 - G_0(z)V][1+G(z)V] = 1$.

\subsection{Single-qubit scattering}

Using the fact that $H$ is box-diagonal in the excitation number space, we will now compute the resolvent $G(E)$ within the single excitation sector. Let us introduce the right-hand-side of Eq.\ \ref{eq:H}
\begin{align}
\ket{\phi} = V\ket{\Phi} =
\left[\sum_s g_{sk_0}^*\sigma^+_s \right]\ket{\Omega},
\end{align}
where $\ket{\Phi}$ was assumed to have a well defined momentum $k_0$ and thus $E^+ = \omega(k_0)+0^+$. We solve Eq.\ \eqref{eq:H} written as, $(E^+ - H) \ket{\psi} = \ket{\phi}$. Grouping terms this leads to
\begin{align}
(E^+ - \Delta_s) c_s - \sum_k g_{sk} \psi_k &= g_{sk_0}^*\\
(E^+ - \omega_k)\psi_k - \sum_s g_{sk}^* d_s &= 0.
\end{align}
The second equation may be readily solved%
giving an equation for $\vc{c}$
\begin{equation}
(E^+-\Delta_s)c_s
 - \sum_{r} \frac{g_{sk}g_{rk}^*}{E^+-\omega_k} c_r
 = g_{sk_0}^*.
\end{equation}
Note that this equation may be written in a much more compact form
\begin{equation}
\vc{c} = (E^+ - \tilde{H})^{-1} \vc{g}_{k_0},
\end{equation}
where the effective Hamiltonian is the one computed in the master equation formalism. For one qubit
\begin{equation}
\tilde{H} = \Delta'
\end{equation}
and the scattering process has a resonance at the renormalized qubit frequency. For two qubits
\begin{equation}
\tilde{H} = \left(\begin{matrix}
\Delta_1' - i\gamma_1/2 & g_{12} + i \gamma_{12}/2\\
g_{12} + i \gamma_{12}/2 & \Delta_2' - i\gamma_2/2\end{matrix}\right)
\end{equation}
and the scattering resonances appear where predicted by the effective two-qubit interaction, broadened by the respective decay rates, $\gamma$ and $\gamma_{12}$.

Note that unlike the Markovian equation, which is approximate, the scattering equations are exact for any separation of the qubits, because they describe the asymptotic states under a stable but sparse flow of individual photons.

\subsection{Relation to transfer matrix models}
\label{sec:status-quo}

The previous subsection shows that the single- and many-qubit resonances are renormalized due to the individual and collective Lamb shifts. This result seemingly contradicts earlier predictions of the scattering models in one-dimensional waveguides and coupled cavities\ \cite{shen05a,shen05b,zhou2008,longo2010}, where no such renormalization is evident. This contradiction is due to the choice of model in those earlier works and has to be analized case by case.

In Refs.\ \cite{shen05a} and \cite{shen05b}, the coupling is written in position space as a boundary condition for two propagating fields. The coupling is local in space at the position of the qubit and the spectra of the photon are continued towards $\omega\to-\infty$, introducing two independent propagating fields that move left- and right-wards. From the point of view of our formalism this results in a constant spectral function $J(\omega)$. Introducing $J(\omega)=J$ in our integrals, together with some cut-offs, we recover that the imaginary part of the kernel is zero, so that $\Delta'=\Delta$, and the individual qubit renormalization is absent.

Another popular model is an array of coupled cavities modeled by a tight-binding model. This model results from applying a rotating-wave approximation in our band model, introducing a discrete set of $N$ modes $\{a_n,a_n^\dag\}_{n=1}^N$ whose Hamiltonian contains a nearest-neighbor hopping term
\begin{equation}
H_{\gamma}= \sum_{n=1}^N \omega a_n^\dag a_n
-J\sum_{\langle n,m\rangle} a^\dag_n a_m.
\end{equation}
Translational invariance allows us to diagonalize the model with a Fourier transform, obtaining
\begin{equation}
\omega_k = \omega - 2J \cos(k),\; g_{sk} = \frac{1}{\sqrt{N}}e^{-ik x_s},
\end{equation}
where now the position $x_s$ is an integer and the coupling once more does not depend on the frequency. Integrating the kernel for this distribution of frequencies gives
\begin{align}
\mathcal{K}(t) &= g^2 \frac{1}{\sqrt{(J-\Delta+\omega)(J+\Delta-\omega)}},
\end{align}
which is real for a qubit inside the band, $|\Delta-\omega|<J$, and purely imaginary outside the band. Thus, in the usual situation in which the qubit may directly interact with the photons, $\Delta'=\Delta$ and we have no Lamb shift.

The lack of single-qubit and collective Lamb shifts in these models is thus a coincidence that arises due to the approximations in the photon field and in the qubit-photon interaction. It is thus not a physically realistic effect and when wishing to do quantitative predictions of experiments we should go back to models such as the ones in Sect.\ \ref{sec:microscopic}

\section{Numerical results}

\begin{figure}
\includegraphics[width=0.8\linewidth]{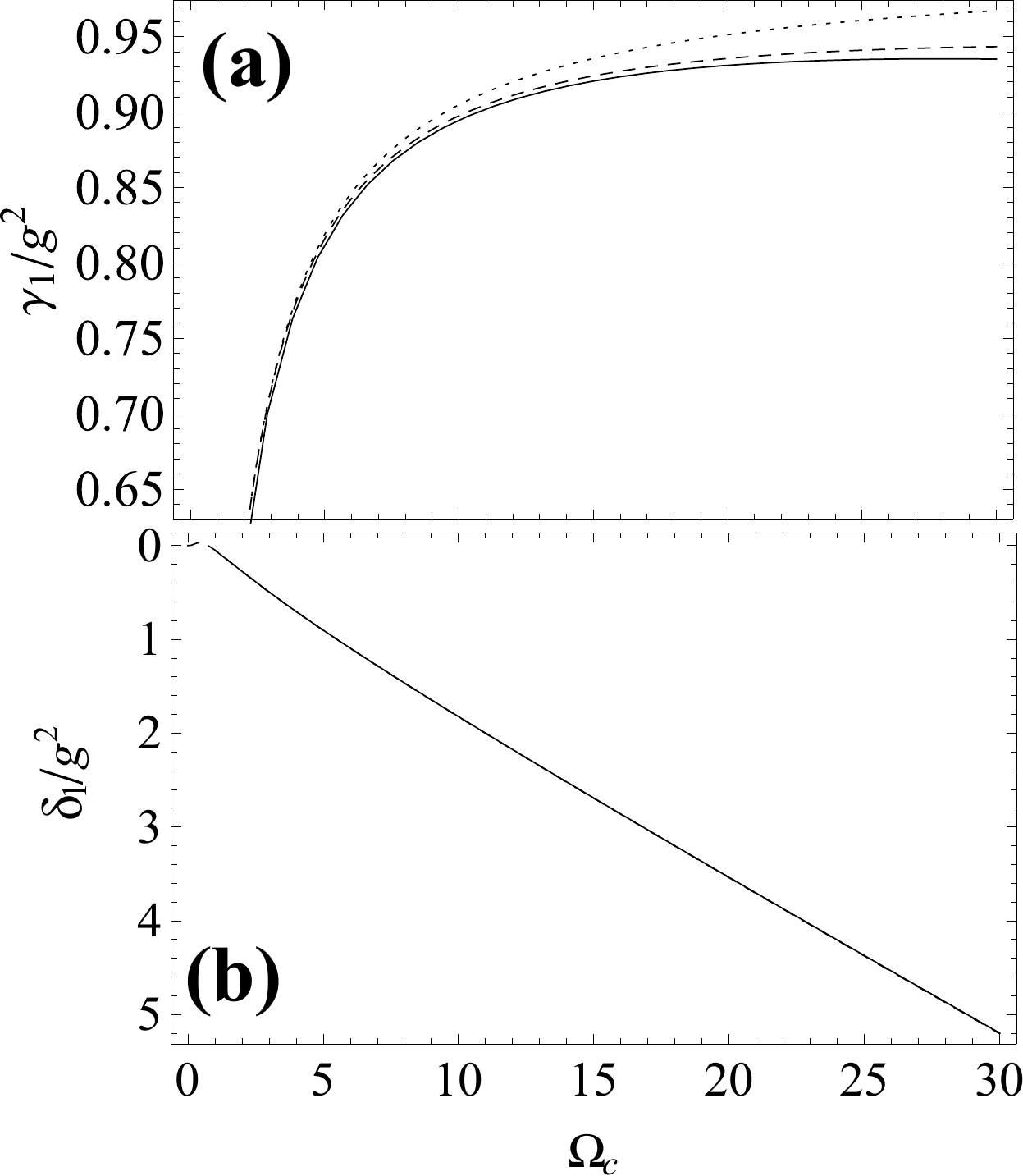}
\caption{Single-qubit spontaneous emission rate, $\gamma_1$, and frequency renormalization, $\delta$, as a function of the model cut-off, $\omega_c$ in \eqref{eq:w-leggett}. We plot the numerical result obtained by solving\ \eqref{eq:integro-differential} with an implicit method (solid), together with the self-consistent expression [\eqref{eq:nqb2},dashed] and the usual prediction without Lamb shift [\eqref{eq:nqb}, dotted].
\label{fig:1qb-exact-numerics}}
\end{figure}

\begin{figure}
\includegraphics[width=0.9\linewidth]{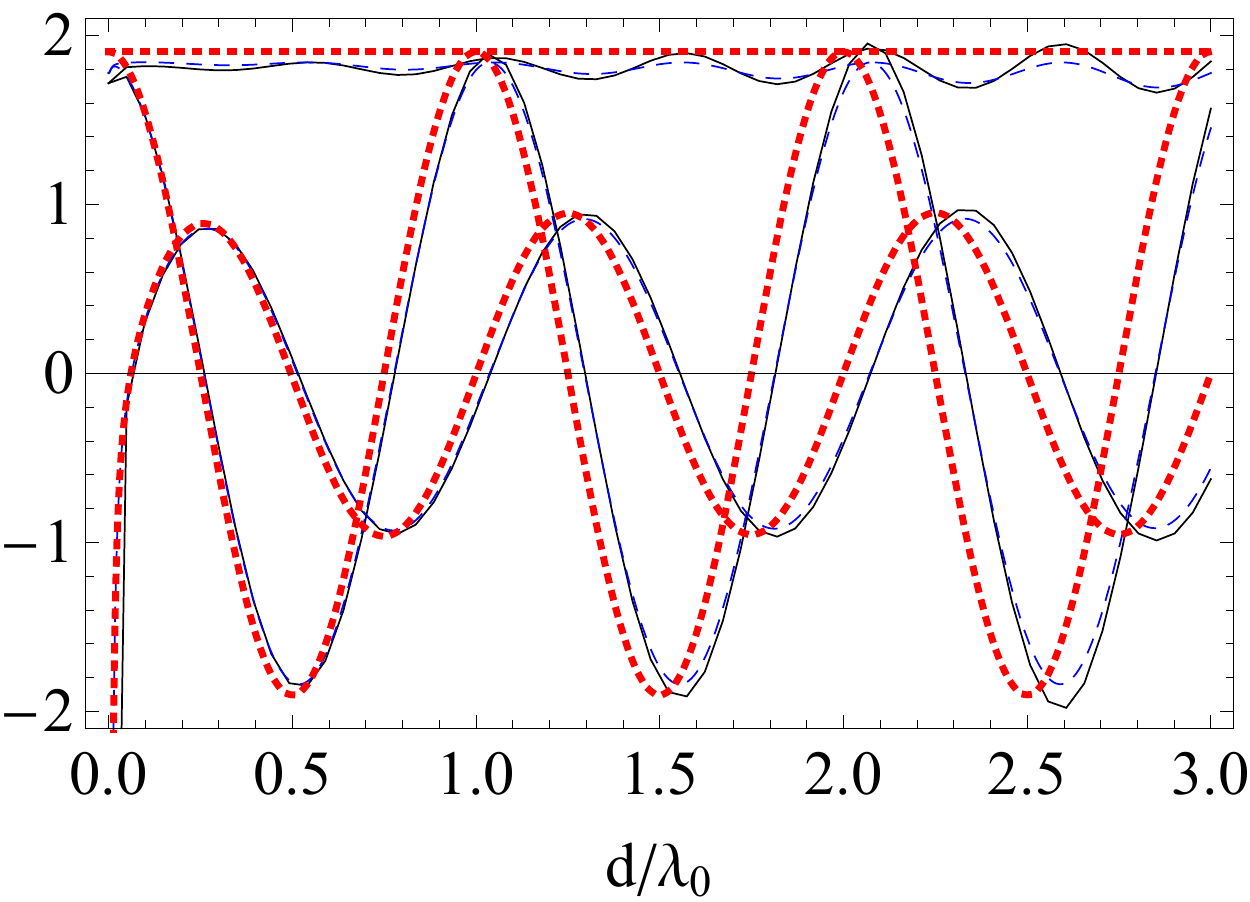}
\caption{Two-qubit parameters, $\gamma$, $\gamma_{12}$ and $g_{12}$, as a function of the model cut-off, $\omega_c$ in \eqref{eq:w-leggett}. We plot the numerical result obtained by solving\ \eqref{eq:integro-differential} with an implicit method (solid), together with our prediction\ \eqref{eq:2qb-parameters} using the self-consistent solution for $\Delta_{\pm}$ (blue, dashed) and expressions for $\gamma,\gamma_{12}$ and $g_{12}$ that do not include the Lamb shifts (red, dotted).\label{fig:2qb-exact-numerics}}
\end{figure}

We now present two numerical approaches to solve the system of two qubits interacting via a transmission line in the RWA. The first set of simulations aim at comparing with our earlier analytic predictions for the toy-model with infinite spectrum\ \eqref{eq:w-leggett}. In this case we solve the integro-differential equation\ \eqref{eq:integro-differential} using the fact that we have explicit expressions for the kernel and that we have an efficient and stable implicit method to solve the equation.

\begin{figure*}[t]
\includegraphics[width=\linewidth]{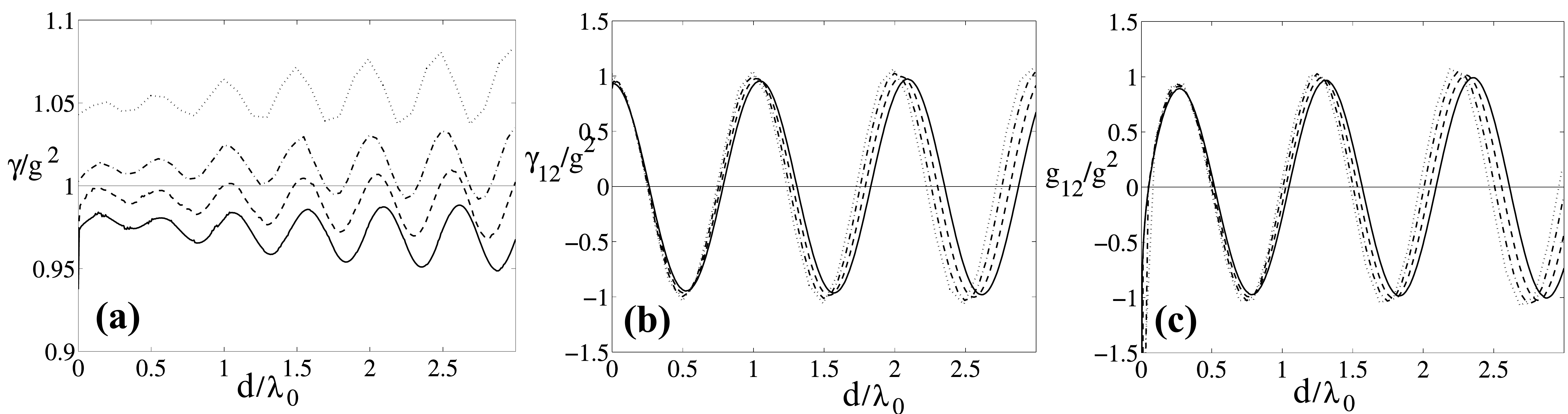}
  \caption{Single-qubit spontaneous emission (a), collective decay (b) and interaction (c) for two qubits interacting with a one-dimensional gapless waveguide, $\omega(k) = \omega_c\sqrt{1-\cos(k\delta x)}$, as a function of the qubit separation $d$. As simulation parameters we use the same gap for both qubits, $\Delta$, and the speed of light $v$ as units. The waveguide has a length $L=20\lambda_0$ and a discretization $\delta{x}=L/N$, with varying number of modes: $N=200$ (dotted), $400$ (dot-dashed), $N=1600$ (dashed) and $N=3200$ (solid).\label{fig:numeric-gapless}}
\end{figure*}

The second set of simulations are more general in their design, working directly with\ \eqref{eq:ode-exact} for a problem with a finite size. This is possible because we work with more realistic scenarios where the spectrum has a physically motivated cut-off, as it is the case of continuous transmission lines\ \eqref{eq:w-gapless} or photonic crystals\ \eqref{eq:w-band}. In both cases we will solve the single excitation model numerically , confirming that many of the features obtained before with the analytically solvable models are present.

The three sections below demand a model fitting mechanism that allows us to extract the parameters $\gamma, \gamma_{12}, g_{12}$ or $\Delta'$ from the dynamics of the qubit, a tool that is also essential to analyze the Matrix Product States simulations in Sect.\ \ref{sec:mps}

\subsection{Exponential cut-off}
\label{sec:implicit}

In this first set of simulations we solve numerically equation\ \eqref{eq:integro-differential}. Our starting point is a formal rewrite of the equation
\begin{equation}
\partial_t c(t) =
-\int_0^t \mathcal{K}(t-\tau)e^{-i\Delta \tau}c(\tau)\mathrm{d}\tau,
\end{equation}
where we have eliminated free evolution, introducing it into the integral. Further integration leaves
\begin{equation}
c(T) = c(0)
-\int_0^T\int_0^t \mathcal{K}(T-\tau)e^{-i\Delta \tau}c(\tau)\mathrm{d}\tau\mathrm{d}t,
\end{equation}
which can also be written
\begin{equation}
c(T) = c(0)-\int_0^T G(T-t)c(t)\mathrm{d}t,
\end{equation}
where the function $G(t)$ can be analytically computed from\ \eqref{eq:w-leggett}, with an expression too large to be reproduced here. The last step is to discretize time
\begin{align}
c(t_n)\simeq c_(t_1) - \frac{\Delta}{2}\sum_{m=1}^{n-1}\big[
G(t_n-t_{m+1})c(t_{m+1}) \nonumber\\
 + G(t_n - t_m)c(t_{m+1})\big].
\end{align}
This equation has been appropriately symmetrized to increase the accuracy and make the formula more stable, but it then implies that the equation must be solved implicitly, as $c(t_n)$ appears also in the sum.

Integration formulas apply directly to the problem of a single qubit's spontaneous emission. In Fig.\ \ref{fig:1qb-exact-numerics} we plot the two parameters as obtained from numerical integration (solid), and from the Markovian formulas, for a problem with a moderate cut-off, $\omega_c=10\Delta$ and coupling strength $g=4\%\Delta$. Notice that while $\delta$ is exactly approximated by the analytical expressions, only the self-consistent renormalization approaches the numerical results.

We can repeat this idea using now two qubits and a slightly larger cut-off, $\omega_c=20\Delta$, to ease plotting. The results are shown in Fig.\ \ref{fig:2qb-exact-numerics}. Notice how the numerical simulations (solid) reproduce the oscillations of $\gamma$ as a function of the distance, which are only captured by our self-consistent formulas (dashed). All other schemes, such as a partial renormalization\ \eqref{eq:simple-renorm2} or the usual resonant dipole approximation with no change in $\Delta$, fail by a significant margin (red dashed lines) that worsens with the distance.

\subsection{Finite length transmission line}
\label{sec:gapless}

\begin{figure}[b]
\includegraphics[width=0.75\linewidth]{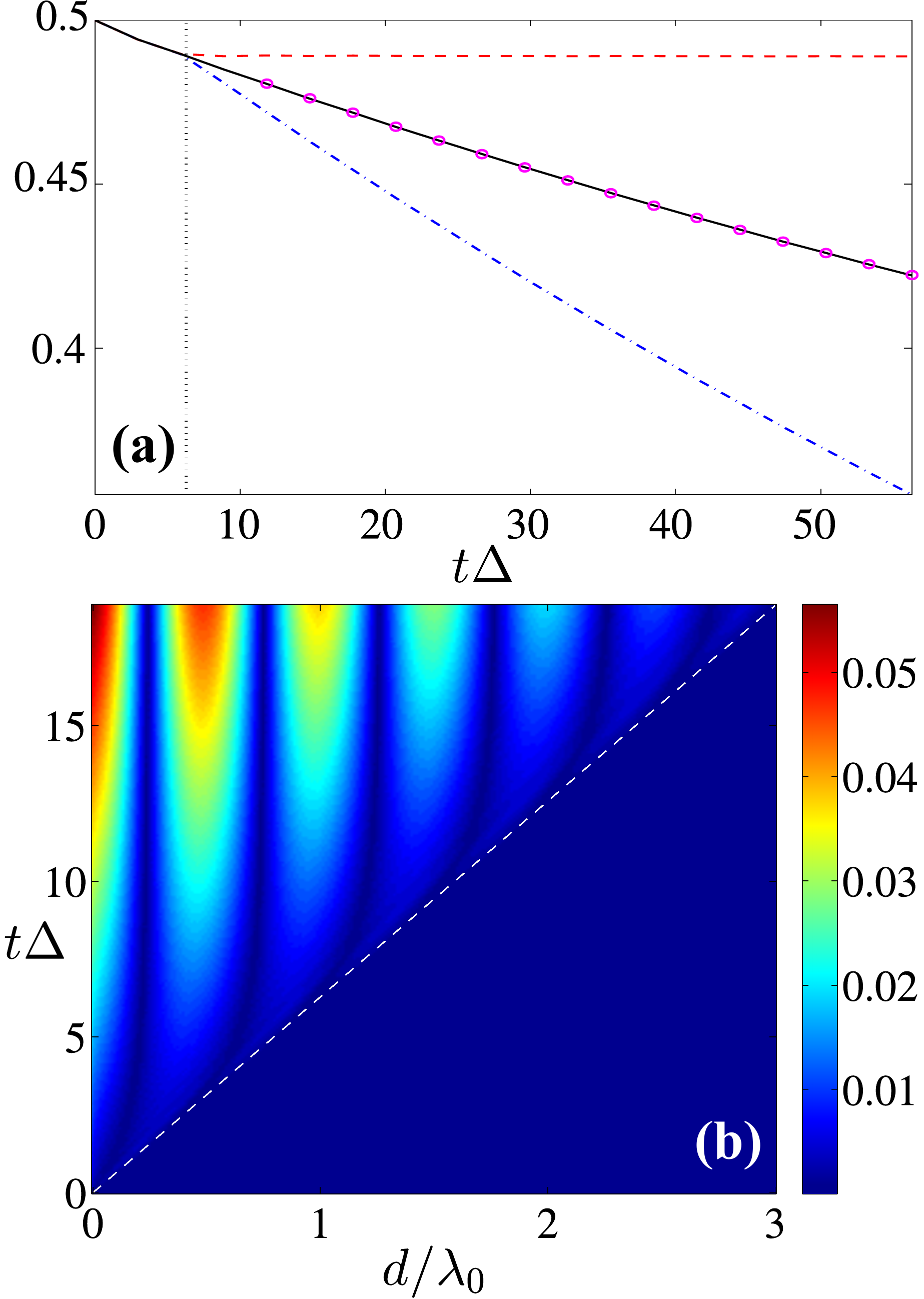}
  \caption{(a) Qubit variables in a single numerical experiment. We plot $|c_+|^2+|c_-|^2$ (solid), $|c_-|^2$ (dashed) and $|c_+|^2$ (dash-dot). Note how $c_+$ and $c_-$ depart at the time at which the photon arrives to the second qubit. The simulation parameters are $L=20\pi$, $N=800$, $g=0.04$ and $d=\lambda_0$. (b) We extend the same plot to all other qubit separations, representing $||c_-|^2-|c_+|^2|$ as a function of time and distance. Outside of the light cone, there is no difference between $|c_-|$ and $|c_+|$, and the qubits are uncorrelated. As guide to the eye we plot the dashed line $d=v t$.\label{fig:fitting800}}
\end{figure}

We now study the first model for which we do not have an analytical solution. This is a discretized model for a one-dimensional transmission line or waveguide\ \cite{peropadre13}. The waveguide will have a total length $L$ divided into segments of size $\Delta{x}$. The coupling between these segments gives rise to the dispersion relation\ \eqref{eq:w-gapless}, which is approximately linear for frequencies smaller than $\omega_c$, the cutoff frequency
\begin{equation}
\omega(k)\approx \omega_c |k| \delta x,
\end{equation}
The speed of light, $v =\frac{\omega}{|k|}=\omega_c \delta x$, will be used together with the qubit frequency $\Delta$ to adimensionalize all quantities: from the unit of length $\frac{v }{\Delta}=1$ we obtain the wavelength of the qubit $\lambda_0=\frac{2\pi v}{\Delta}=2 \pi$, and also the cutoff frequency $\omega_c=\frac{ v }{\delta x}$.

\begin{figure}[t]
\includegraphics[width=0.7\linewidth]{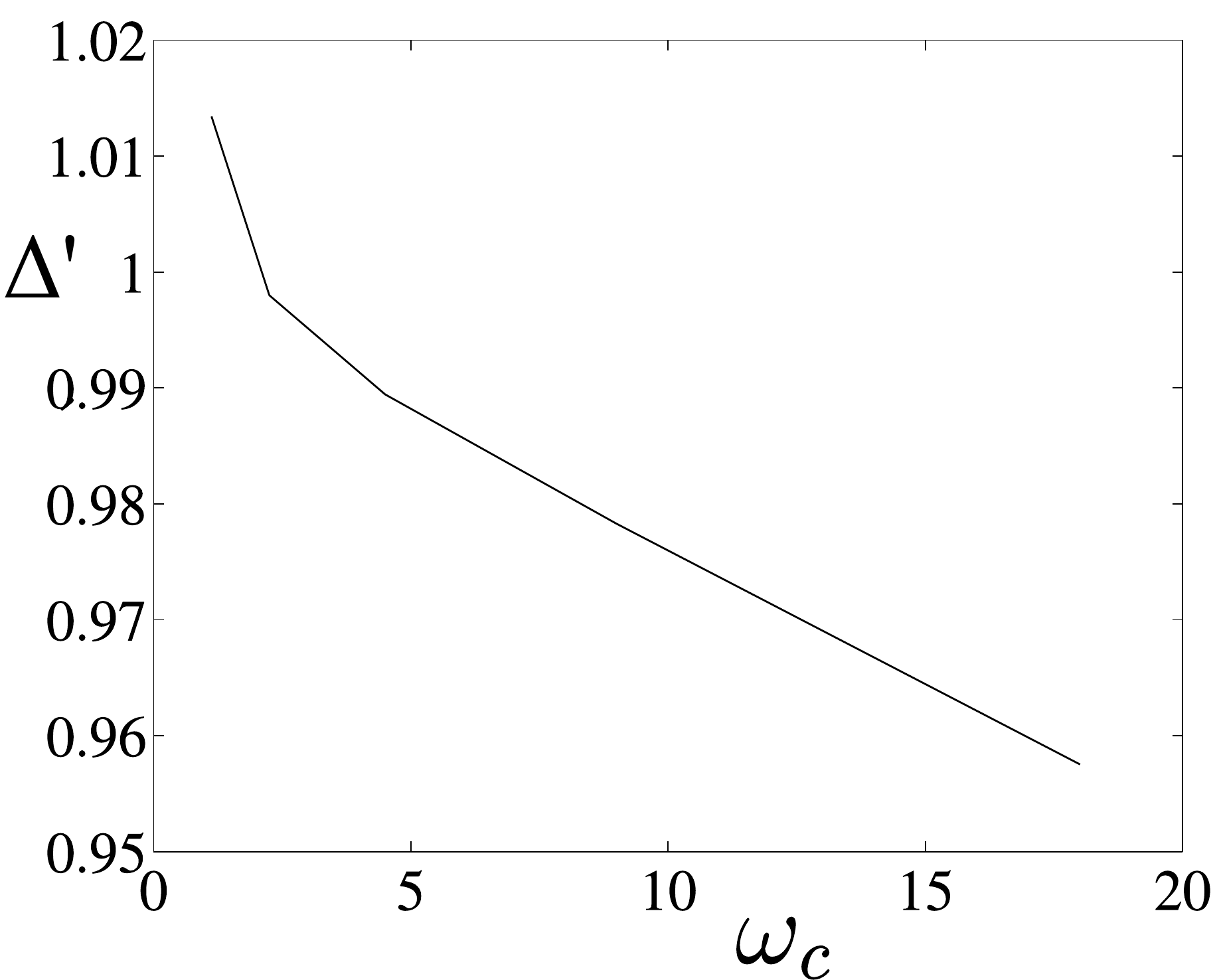}
  \caption{Renormalized gap of the first qubit obtained from the approximate period of the functions $g_{12}$ and $\gamma_{12}$, as a function of the cut-off $\omega_c$, for $\Delta=1$, $g=4\%\Delta$.\label{fig:renormD}}
\end{figure}

For the simulation to actually reproduce propagating photons, the waveguide length $L$ also should be larger than the qubit wavelength and the total simulation time should be shorter than the time for a photon to cross the whole line and bounce back (open boundary conditions) or return the qubit from the other side (periodic boundary conditions). In our simulations we take $L$ between $10 \lambda_0$ and $20 \lambda_0$ and thus $T< L/ v $ to avoid the revivals.

From the dispersion relation and the modes, we solve numerically the full equation in the single-excitation limit\ \eqref{eq:ode-exact}. This is done with MATLAB's adaptive ODE solvers using an initial condition in which only one qubit is excited and the line has no photons. The excited qubit will relax and spontaneously emit a single photon which, after a finite travel time $t_{flight}=d/ v $ through the separation $d$, is scattered by the second qubit. It is important to remark the existence of an approximate light cone. Outside this light cone, $t<t_{flight}$, the first qubit does not have enough time to significantly influence its unexcited counterpart and thus the Markovian model does not apply. After the time of flight time $t_{flight}$, we may start talking about interaction and the collective variables $c_{\pm}$ start having independent dynamics.

In Fig.\ \ref{fig:fitting800}a we show all dynamical variables from one particular simulation. First of all, notice how $|c_1|^2+|c_2|^2=|c_+|^2+|c_-|^2$ follows an exponential decay with a rate, $\gamma$, that remains approximately constant throughout the whole evolution. At $t_{flight}$ the traveling photon hits the unexcited qubit, which partially absorbs it. At this point, variables $c_+$ and $c_-$ depart and acquire different exponential decay rates, whose average is $\gamma$ and whose differece is $\gamma_{12}$. Finally, in Fig.\ \ref{fig:fitting800}b we plot the value of $|c_+|^2-|c_-|^2$ as a function of time, thereby showing how the light cone extends approximately over all distances and how at long times the dynamics is periodic on the qubit separation.

\begin{figure*}[t]
\includegraphics[width=\linewidth]{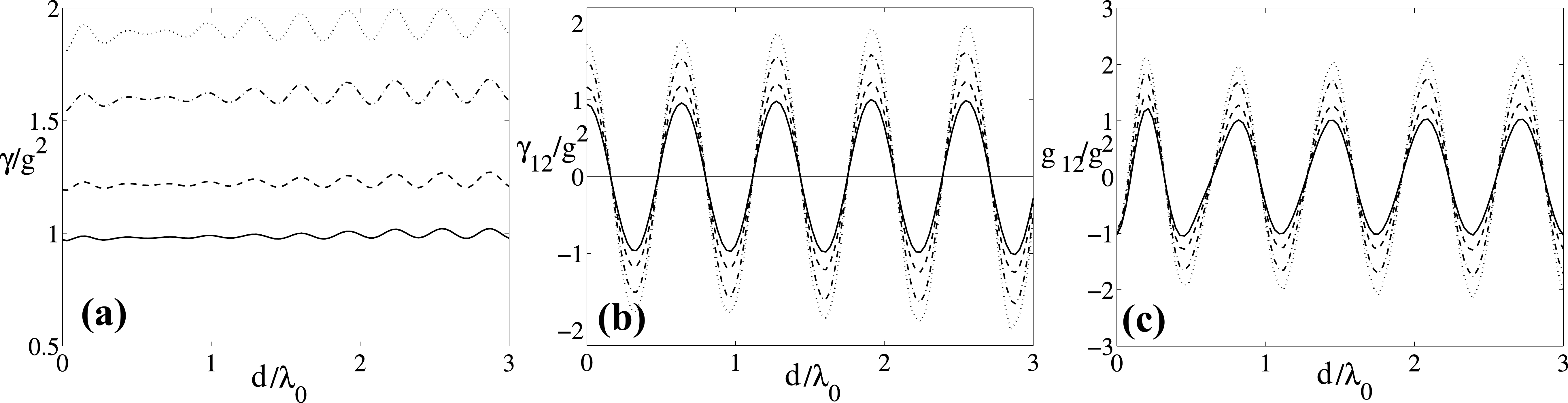}
  \caption{Single-qubit spontaneous emission (a), collective decay (b) and interaction (c) for two qubits interacting with a one-dimensional gphotonic crystal, $\omega(k) = \omega_0-J\cos(k)$. We use periodic boundary conditions for qubits with gap $\Delta=\omega_0$, $N=800$, a waveguide length $L=20\lambda_0$ and a varying bandwith $J/\omega_0=0.5$ (dotted), $0.6$ (dashed), $0.8$ (dot-dashed) and $1$ (solid). Note how the period of the curves is not significantly affected by the bandwidth, $J$. \label{fig:numeric-band}}
\end{figure*}

We can simulate multiple different processes to recover the dependency of $\gamma$, $\gamma_{12}$ and $g_{12}$ on the qubit separation $d$. This is shown in figure Fig.\ \ref{fig:numeric-gapless} for different discretizations of the chain. In each curve we employed a different number of sites $N=200,400,1600,3200$, but always the same length $L=20 \lambda_0$. Note that both $g_{12}$ and $\gamma_{12}$ are approximately periodic, as in the analytically solvable models, with a period that depends on the discretization and thus on the cut-off frequency. This is a sign of the qubit gap renormalization.

Studying the photon wavelength from the zeroes of the $\gamma_{12}$ and $g_{12}$ we are able to obtain the dependance of $\Delta'$ as a function of $\delta x$ or $\omega_c$ as it is showed in Fig.\ \ref{fig:renormD}. The value of the individual spontaneous emission rate $\gamma$ was also predicted to oscillate due to the influence of the collective Lamb shifts in the decay rate. When we average over $\gamma(d)$ over the qubit separation, this collective Lamb shift is averaged out and we obtain a mean value that only depends on $\Delta'$ and, since $J(\Delta')\propto\Delta'$, follows a similar curve to Fig.\ \ref{fig:renormD}.

\subsection{Photonic crystal}
\label{sec:band}

In this second case instead of having a transmission line with free travelling photons, we consider a photonic crystal\ \eqref{eq:w-band}. This dispersion relation may be obtained, for instance, by coupling together multiple cavities or patterning a waveguide. In this case the photon's speed, determined by the group velocity, depends on its frequency. We will fix $\omega_0=\Delta$, so that this group velocity is strictly the coupling strength $J$
\begin{equation}
v_g = \left.\frac{\partial\omega_k}{\partial k}\right|_{\omega_k=\Delta} = J
\end{equation}
The actual length of the photonic medium is $L=N$ the number of sites in the crystal. However now the dispersion relation does not depend on $N$ and the number of modes will only influence features such as the propagation time of the photons and the smoothness of the curves.

As before, the numerical integration works with the full model\ \eqref{eq:ode-exact}, now studying only the influence of the band width $J$ and the qubit separation $d$. Once more, we find an approximate light cone, this time governed by the group velocity $v_g$.  Once more we compute the collective variables $c_{\pm}$ numerically inside and outside the light cone and fit the numerical results to the theoretical curves.

In Fig.\ \ref{fig:numeric-band} we plot the fitted parameters  $g_12$, $\gamma$ and $\gamma_{12}$ as a function of the qubit separation for different bandwidths, $J$. Note how the spontaneous emission rate and the interactions grow with decreasing bandwith. Physically this may be understood as a consequence of the reduced photon velocity and an increase in the spectral density around the two-level system, which facilitate the interaction between the second qubit and the photon that was originally emitted. Note also that due to the very small cut-off, which cannot grow beyond $2\Delta$, there is a negligible renormalization of the qubit frequency, which is almost invisible in $\gamma$ and impossible to recover from the periods of $\gamma_{12}$ and $g_{12}$.

\section{Matrix Product States}
\label{sec:mps}

We finish the theoretical study by performing the same simulations in a more general situation, that is the full spin-boson model\ \eqref{eq:spin-boson} without resorting to the single-excitation or RWA approximation. We do this by using the Matrix Product State ansatz for solving numerically the full spin-boson model with both qubits interacting with the bosonic bath. The method used is a generalization of the one outlined in\ \cite{peropadre13}, introducing one more qubit and applying the same model fitting techniques to extract the parameters $\gamma,\gamma_{12}, g_{12}$ and $\Delta'$.

\begin{figure}
\includegraphics[width=0.8\linewidth]{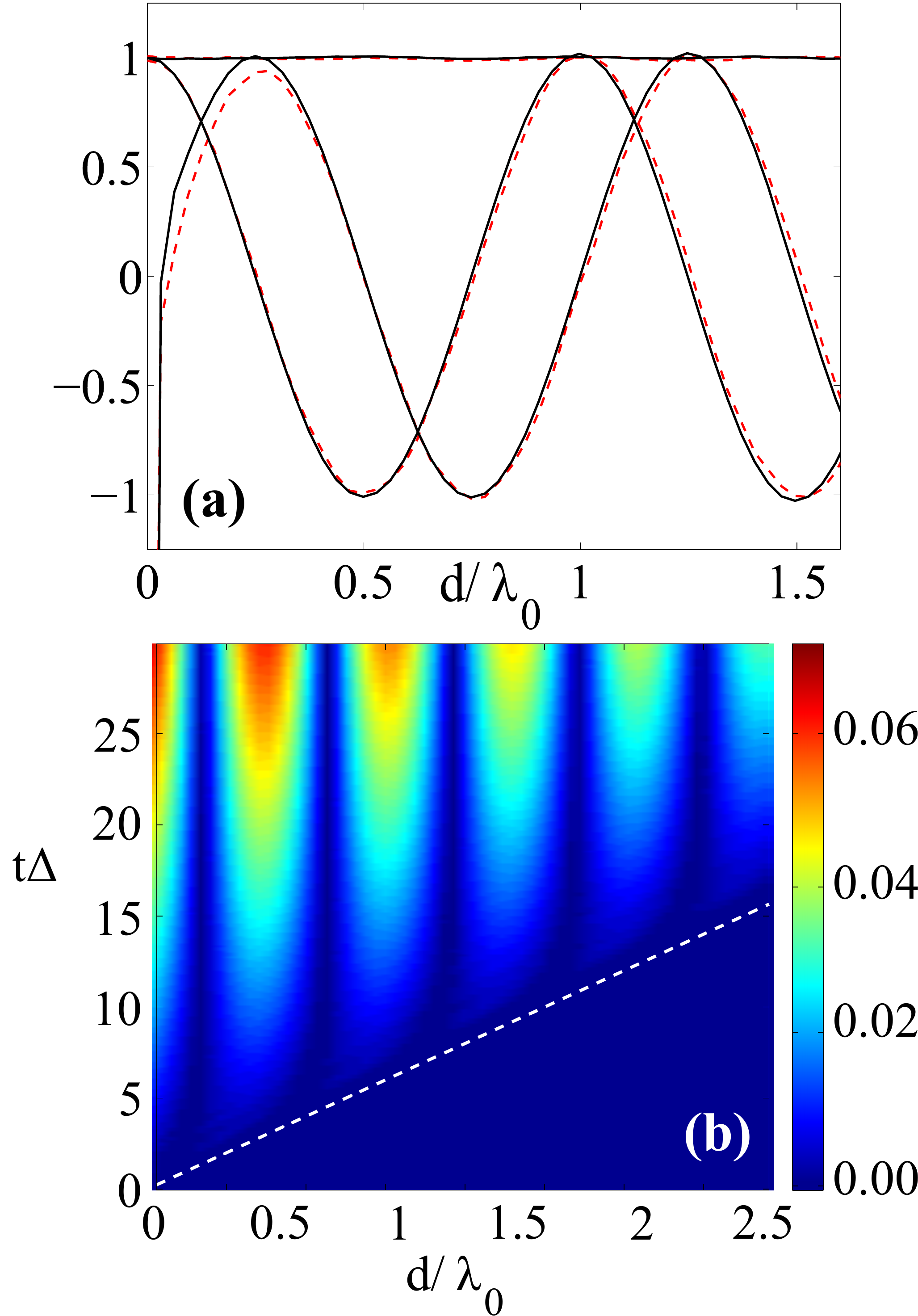}
  \caption{(a) Comparison between the results of MPS simulations, $\gamma/2g^2$, $\gamma_{12}/2g^2$ and $g_{12}/g^2$, in solid line, and the RWA method, in dashed line. (b) Difference $||c_+|^2-|c_-|^2|$ as a function of time and qubit separation, together with the light-cone (dashed). The parameters of the simulation are $N=321$, $g=0.05$, $\Delta=1$ and $ v =1$.\label{fig:MPS}}
\end{figure}

The scope of this paper the study of qubit-qubit interactions for realistic couplings, with $g\le 5\%\Delta$, such as the ones describing transmon or ordinary flux qubits close to an open transmission line. The outcome of these simulations is shown in Fig.\ \ref{fig:MPS}, where we plot the Markovian parameters from one MPS simulation together with a numerical simulation using the RWA and single-excitation limit.

The main message from those simulations is therefore that the RWA accurately describes the qubit and photon dynamics for the weak and strong-coupling regimes. The numerical and theoretical results from earlier sections should therefore apply to ongoing experiments and could be verified in state-of-the art setups that already study photon-qubit interactions.

\section{Summary and conclusions}
\label{sec:summary}

We have studied the dynamics of one and two qubits coupled to an open transmission line. We have related the dynamics of these qubits, in a regime of weak or strong coupling $(g/\Delta<5\%)$ to the RWA or single-excitation equations. From these equations we have derived carefully the Markovian limit, studying the approximations involved and paying special attention to the single-qubit and collective Lamb shifts that arise. We showed how the Markovian parameters may carry a significant and measurable dependence on the microscopic details of the underlying photons, such as their ultraviolet cut-offs and the exact shape of the spectral function. Such features are present not only on the non-equilibrium dynamics of the qubits, but also on the spectroscopy or scattering properties. Our predictions have been confirmed for a large variety of physical models, and also with numerical simulations based on Matrix Product States without RWA approximations.

One take-home message of this work is that the microscopic details of qubits interacting with propagating photons can have a measurable impact in the quantum-optical properties of the qubits. In order to experimentally probe those effects one needs to be able to change such microscopic models. As shown in Ref.\ \cite{garcia-ripoll2014}, a very useful setup to realize such changes has already been demonstrated in the lab\ \cite{shanks13}. The experiment would consist on mobile transmon qubits that are suspended over a transmission line, thereby allowing us to change the coupling strength and the qubit-qubit separation, and thus obtaining not only the curves $\Delta', \gamma, \gamma_{12}$ and $g_{12}$, but also the bare parameters in the absence of renormalizations.

The authors acknowledge support from Spanish Mineco project FIS2012-33022, CAM Research Consortium QUITEMAD+ (S2013/ICE-2801), EU Marie Curie C.I.G. 630955 NewFQS, and from EU FP7 FET-Open project PROMISCE. JJGR acknowledges computer resources and assistance provided by the Centro de Supercomputacion y Visualizacion de Madrid (CeSViMa) and the Spanish Supercomputing Network.

\appendix

\section{Master equation constants}
\label{app:kernel}

We have related the dynamics of spontaneous emission to a linear system of differential equations that is local in time. In this setting, the coupling between spins and the decay of the excited populations is given by a matrix
\begin{align}
\bar{H}_{ss'} &= - i\int_0^\infty K_{ss'}(\tau)
e^{i\Delta_{s'}'\tau}\mathrm{d}\tau,\\
&= (\Delta_{s}'-\Delta_s -i\gamma_{s})\delta_{ss'} +
	(g_{ss'} - i\gamma_{ss'})(1-\delta_{ss'})\nonumber
\end{align}
In order to relate this expression to the scattering equations we will focus on the time integral, expressing it in the original couplings as the limit
\begin{align}
\bar{H}_{ss'} & = \lim_{\varepsilon \to 0^+} -i\int_0^\infty
\sum_k g_{sk}g_{s'k}^* e^{-i\omega_k\tau+i(\Delta_{s'}'+i\varepsilon)\tau}
\mathrm{d}\tau\nonumber \\
& = \lim_{\varepsilon\to0^+}
\sum_k \frac{g_{sk}g_{s'k}^*}{i(\Delta_{s'}-\omega_k)-\varepsilon}
\left[e^{[i(\Delta_{s'}-\omega_k)-\varepsilon]t}\right]_0^\infty
\nonumber\\
& = \sum_k \frac{-ig_{sk}g_{s'k}^*}{\Delta_{s'}'+i0^+-\omega_k}
\end{align}
Note how this is close to the expression that appears in the denominator of the Lippmann-Schwinger equation: unlike the master equation, the LS formalism deals with stationary states, a situation in which all amplitudes $c_s$ and $\psi_k$ evolve with the same frequency. In that case we have to replace $\Delta_{s}'\to E$, and we recover the effective Hamiltonian from the scattering states.

%

\end{document}